# MUREN: delivering edge services in joint SDN-SDR multi-radio nodes

Mathieu Bouet, Vania Conan, Hicham Khalifé, Kévin Phemius, Jawad Seddar
Thales Communications & Security
{name.surname}@thalesgroup.com

*Abstract*—To meet the growing local and distributed computing needs, the cloud is now descending to the network edge and sometimes to user equipments. This approach aims at distributing computing, data processing, and networking services closer to the end users. Instead of concentrating data and computation in a small number of large clouds, many edge systems are envisioned to be deployed close to the end users or where computing and intelligent networking can best meet user needs. In this paper, we go further converging such massively distributed computing systems with multiple radio accesses. We propose an architecture called MUREN (Multi-Radio Edge Node) for managing traffic in future mobile edge networks. Our solution is based on the Mobile Edge Cloud (MEC) architecture and its close interaction with Software Defined Networking (SDN), the whole jointly interacting with Software-Defined Radios (SDR). We have implemented our architecture in a proof of concept and tested it with two edge scenarios. Our experiments show that centralizing the intelligence in the MEC allows to guarantee the requirements of the edge services either by adapting the waveform parameters, or through changing the radio interface or even by reconfiguring the applications. More generally, the best decision can be seen as the optimal reaction to the wireless links variations.

## I. Introduction

The recently launched 5G research programs have identified a set of new requirements and objectives [1]. These new generation networks ambition to offer higher rates and capacities to end users, cope with wireless networks densification with the spread of new devices and access technologies, guarantee lower delays for remotely controlling devices, but also offer cost reduction through an easy to configure fully automated and decentralized control plane.

For these reasons, Mobile Edge Computing (MEC) has recently emerged as a key approach towards 5G [2]. An industry specification group of the same name was created in the fall of 2014 at the European Telecommunications Standards Institute (ETSI). It aims at providing content providers with cloud-computing capabilities and an IT service environment at the edge of the mobile network, within the Radio Access Network (RAN) and in close proximity to mobile subscribers. The main objectives are to reduce latency, ensure highly efficient network operation and service delivery, and offer an improved user experience. MEC is based on a virtualization platform that leverages Network Functions Virtualization (NFV) and Software-Defined Networking (SDN) to run applications at the edge of the network.

In this paper, we go further in the MEC approach. We address the challenge of using in real-time radio network information to improve service operation and thus user experience. To this end, we combine Software-Defined Radios (SDR) and SDN on a MEC platform to jointly use radio and virtualization resources. The generic framework we propose is named MUREN for *Multi-Radio Edge Node*. It supports multi-homing and multi-RAT (Radio Access Technologies) at the edge. We implemented a prototype of MUREN on a real testbed. The SDR part is based on USRP and GNU Radios while the SDN/NFV part is based on Open vSwitch and light containers. We evaluate our approach with two scenarios showing the capacity to: i) reconfigure radios in an agile way to meet application needs and ii) locally and flexibly re-chain and reconfigure applications when radio resources are not sufficient.

The rest of this paper is organized as follows. First, Section II presents the background and state of the art. Then, Section III presents the MUREN architecture composed of a virtualization platform and Software-Defined radios. Our implementation of MUREN, the different scenarios we considered and their evaluation are detailed in Section IV. Finally, Section V concludes this paper.

## II. Background and Related work

In this section, we present a comprehensive background on the three pillars of our approach.

### A. Mobile Edge Computing (MEC)

MEC is currently specified at the European Telecommunications Standards Institute (ETSI) [2]. It has recently emerged as a new evolution of mobile networks. MEC proposes to leverage the cloud computing and SDN approaches to deploy various applications and content caching on cloud computing-like capabilities at the edge of the networks. The rationale behind this approach is that, by bringing applications and resource-heavy tasks closer to the users, i) network congestion can be minimized, ii) the latency between the users and the applications can be reduced, and iii) operators can optimize their infrastructures and differentiate their services. For example, instead of hosting a gaming service in a centralized datacenter far from the user, the MEC approach will allow to deploy this application at the edge, close to the user, thus reducing bandwidth and latency and increasing quality of experience.

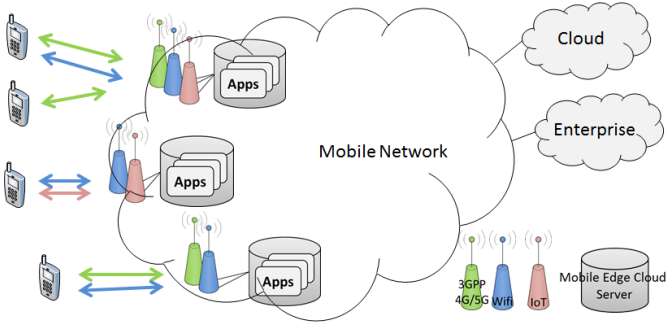

Fig. 1: MEC context.

The concept of edge cloud/computing is not new. For example, cloudlets were envisioned for hostile environments and tactical networks [3] and fog computing for IoT services [4]. However, the MEC initiative constitutes the first formalized and industry-driven framework. The first release of the MEC architecture has been recently issued by the dedicated working group of ETSI [5].

*B. Software-Defined Networking (SDN)*

Software-Defined Networking (SDN) has emerged as an efficient approach to make network management and service deployment more flexible and more programmable. It aims at automating network configuration and facilitating the support of network services in virtualization environments via a logical separation of the data, control and management planes [6], [7]. For example, the OpenFlow protocol was one of the first SDN protocols to specify the access to the forwarding plane of switches [8]. It relies on at least one SDN controller that uses this communication protocol to enforce flow forwarding/routing in a software-defined network. Such a flow granularity enables to have specific routing and traffic engineering policies either depending on the source, the destination and/or the application.

First adopted in datacenters, SDN is now considered for WAN, mobile networks and wireless networks. A comprehensive survey on SDN has been proposed by [9]. Very recently, [10] discussed general schemes illustrating how SDN can bring intelligence in 5G heterogeneous networks while [11] proposes a cellular SDN framework and [12] a SDN-based LTE core network. SoftAir [13] investigated wireless base station decomposition using the SDN approach. We go further on the convergence integrating SDN and SDR in a MEC approach.

*C. Multiple and flexible Radio Devices*

Most off the shelf radio devices today are equipped with multiple wireless interfaces. Indeed, any mobile phone can operate with 2G/3G (even 4G) standards in addition to supporting WiFi and bluetooth technologies. These interfaces are often managed independently or bridged at the transport layer [14] considering mainly the availability of a single interface at a time.

In parellel, the wireless research community has recently deployed considerable efforts to develop flexible radios capable of dynamically adjust their parameters in order to better exploit the changing wireless environment. In fact, a new generation of smart, programmable radios has emerged enabled by the software defined radio (SDR) paradigm and empowered by interference sensing, environment learning, and dynamic spectrum access through the cognitive radio research [15]. However, this agility was often considered over a single radio interface.

More recently, the emergence of COTS based on SDR allows to easily implement multiple radio technologies on the same radio device. Moreover, some open source initiatives such as OpenBTS [16], and OpenAirInterface [17] for LTE prototyping make integrating and testing such solutions possible with relatively limited efforts. Nevertheless, the challenge now is in selecting the appropriate waveform for the right traffic based on the required QoS guarantee and also exploiting the system as efficiently as possible. For instance, one would schedule low data rate SMS type messages on highly reliable low data rate interface and video streaming requiring less reliability on the LTE interface. More challenging is undergoing online adaptation (without service interruption) on the radio link for additional video admission for example as enabled by the SDR paradigm.

## III. THE MUREN ARCHITECTURE

The MUREN architecture framework follows the guidelines of the ETSI MEC framework presented in [5], and more precisely the so called mobile edge host level.

A MUREN node is thus an entity that contains the mobile edge platform and a virtualization infrastructure which provides compute, storage, and network resources for the mobile edge applications. The virtualization infrastructure includes a data plane, based in our case on SDN, that executes the traffic rules received by the mobile edge platform, and forward the traffic among applications, services and networks (3GPP, WiFi, IoT...). As depicted on Figure 2, a MUREN node is composed of four parts that interact:

- **The Waveform/Radio part**, which is not fully considered in the MEC specifications yet, is composed of multiple programmable radios (or SDRs) managed by an SDR controller. These radios constitute in fact the transceivers where a number of waveforms are implemented allowing the device to operate on various spectrum bands using different access techniques. For timely and accurate assessment of the wireless environment, the waveform part constantly measures and collects Channel State Information (CSI) on each available waveform. The CSI includes packet error rate, received signal strength, estimated radio throughput, queue size, user equipment contexts etc. These well-known measures are gathered, in our architecture, by the SDR controller, which exposes them to the MEC server. Upon decision of the MEC controller, each waveform can be independently reconfigured.
- **The Virtualization Platform part** hosts mobile edge applications and services using virtualization. It is also

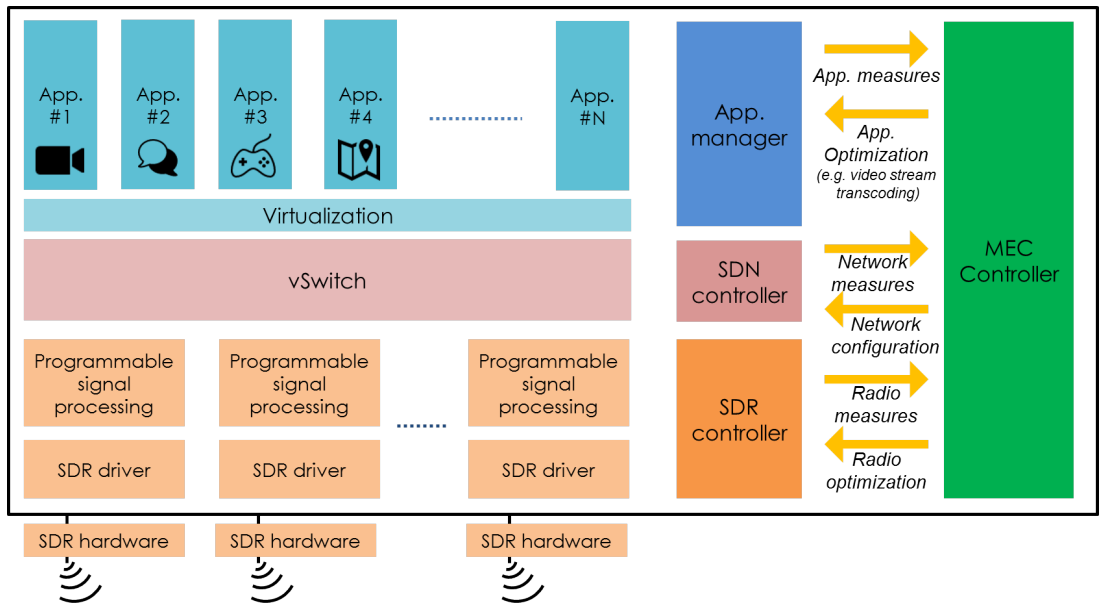

Fig. 2: The MUREN architecture.

in charge of steering and forwarding the traffic from/to the applications and the radio interfaces in the correct sequence, thus instantiating the service chain decided by the MEC controller. This is supported by a vSwitch (an SDN switch) configured by an SDN controller. Using the vSwitch, the SDN controller also monitors the network interface of each application. It measures the throughput, the packet error rate and potentially the jitter using the vSwitch. These information are exposed to the MEC controller.

- **The Application part** is where the applications run in the MEC edge host. Each application can be viewed as a single purpose block with a well defined function that is run in general as Virtual Machine (VM) or container on top of the virtualization layer. The applications can have a number of rules and requirements associated to them, such as required resources (compute, storage), maximum latency, encoding bitrate etc. The App Manager is in charge of instantiating and configuring the applications and collecting and reporting performance and fault information. It exposes these information to and get order from the MEC controller.
- **The MEC controller part** is the brain of the edge host. Its goal is to satisfy application SLAs according to available resources both at the virtualization layer (compute, storage, network) and at the waveform/radio layer. For this reason, it interfaces with the SDR controller, the SDN controller and the App Manager. Based on monitoring information from these three layers, it decides to act either on the programmable radios (e.g. enabling Forward Error Correction - FEC), on the platform (e.g. re-routing flows to another radio) or on the applications (e.g. changing parameters, instantiating a new application etc.). For example, when the QoS of a multimedia stream is not reached or degrades, depending on the causes, the MEC controller might decide to i) change the bitrate of the application via the App manager, ii) change the bit rate of the stream via the App manager which will instantiate a new encoding application and via the SDN controller which will steer the traffic from the multimedia application to the encoder and from the encoder to the network interface, iii) switch the stream an another wireless interface via the SDN controller, iv) reconfigure the concerned waveform (for example activate FEC) via the SDR controller, or v) a combination of these actions.

## IV. PROOF OF CONCEPT IMPLEMENTATION AND EXPERIMENTATION

### A. Implementation and testbed

In order to prove the efficiency of MUREN and evaluate its capability to address the specific challenges of mobile edge computing, we have implemented the proposed solution in a validation testbed. The hence obtained proof of concept emulates the edge nodes of Figure 1 while abstracting the backbone and central clouds. Indeed, we focus on the portion of the network where our contributions lie.

In our implementation of the waveform part of the MUREN architecture, we used USRP radios from Ettus research [18]. These SDR equipments can be used with a variety of software libraries to create complex applications. We chose to go with GNU Radio [19] as we have some previous experience with the tool. Furthermore, it contains quite a few essential building blocks readily available.

We emulated a multi-radio system by using two USRP N210 devices and bridging them together on a PC. This provides the computer with two distinct virtual wireless interfaces with IP capabilities. Depending on the scenario, the radio links have different properties. For all scenarii, our URSP devices are equipped with two antennas (1 TX and 1 RX). We use Frequency Division Duplex (FDD) and GMSK modulation with different bandwidths. The first interface is a *High Rate (HR)* interface while the second is a *Low Rate (LR)* interface.

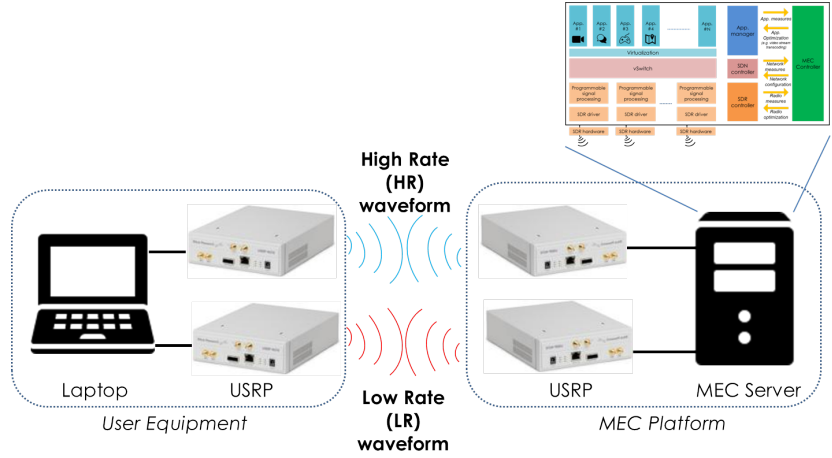

Fig. 3: Our experimental testbed is composed of our MUREN implementation hosted on a PC which acts as a MEC server and communicates with a user equipment-like PC through two pairs of USRPs.

Several low layer indicators were made available by our SDR controller to our MEC controller, such as load on any of the links or Packet Error Rate (PER). There are also several actions that can be taken by the MEC controller and enforced by the SDR controller such as changing the bandwidth, the operating frequency, the code rate etc.

The Virtualization platform is composed of a custom SDN controller that gathers monitoring information from the network interfaces and enforces the policies decided at the MEC controller. The service chaining is assured by an Open vSwitch bridge [20], managed by the SDN controller through the OpenFlow protocol [8].

The SDN controller constitutes an additional degree of freedom (i.e a second level of reaction) to guarantee application requirements. For instance, the SDN controller can steer an application's traffic from the HR interface to the LR interface if additional reliability (and coverage) is required for low rate applications. The MEC controller can impose such decisions when the possible actions at the waveform layer, such as adapting modulation and coding schemes or changing code rate, are deemed inefficient or unachievable.

To implement the Application part, we refrained from using full-fledged virtual machines to host our applications. We instead used containers based on Linux Network Namespaces. This lightweight implementation was more suited to our experimentation with the added gain of rapid instantiation of applications. The App manager is a small program capable of instantiating a Network Namespace connected to the vSwitch and running a single process (e.g. a video server, a video transcoder, a video watermarker etc.). These containers can be brought up and down quickly depending on the need of the service chain.

A key aspect to point out is that an individual application is unaware of the service chain. Only the SDN controller has the knowledge to instruct the vSwitch to forward the traffic. The vSwitch can also modify the headers of the packets if needed to ensure that the traffic will flow correctly through the containers so that everything is transparent to the applications.

Note that to simplify the evaluation and master the experimental conditions, a human operator plays the role of the MEC controller in our testing platform. In other words, the decisions are taken manually in our experiments. This process could easily be automatically driven by specified policies defined at the MEC controller and thus emulates a decision engine. However, these policies will be highly dependant on the targeted services and their QoS requirements. Such a smart automation is part of future work.

TABLE I: Defined SLAs by application type (Scenario 1).

| Application | Requirements | Measured KPIs |
|---|---|---|
| SMS | Bounded delay and high reliability | RTT $\leq$ 50ms and 99.99% success rate |
| Video1 | Good QoE[1] | Bounded load on the HR waveform |
| Video2 | Good QoE | Bounded load on the LR waveform |

Our different experiments rely on two computers, each of them attached to two USRP devices. The first computer plays the role of a MEC server co-located with a Radio Network element while the second one is a User Equipment (UE).

### B. Scenario 1: Guaranteeing the delay of critical SMS

In this scenario, we consider that the HR waveform has a theoretical raw capacity of 1 Mbps while the LR waveform can reach a maximum of 125 kbps. We also presume that the HR interface, while providing a higher capacity, is less reliable than the LR interface. Such a configuration can be considered realistic in wireless environments where the higher bands offer considerably higher data rates at the price of lower coverage distances and reduced resilience to distortions.

In our test, three different applications are considered with traffic sent from the MEC infrastructure to the UE:

- A messaging application, with strong RTT and reliability requirements. This can be a critical SMS application (e.g. critical IoT application).
- A high bitrate video application (Video1).
- A lower bitrate video application (Video2).

Based on these applications, we define several SLAs that the MEC operator must fulfill. In this instance the SLAs are highlighted in Table I.

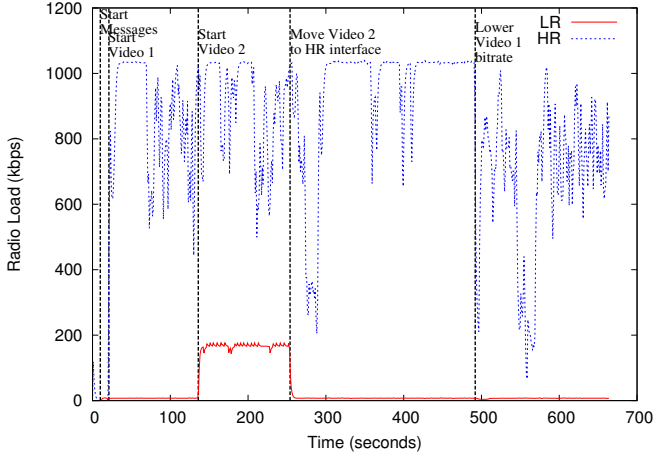

Fig. 4: Waveform loads for Scenario 1.

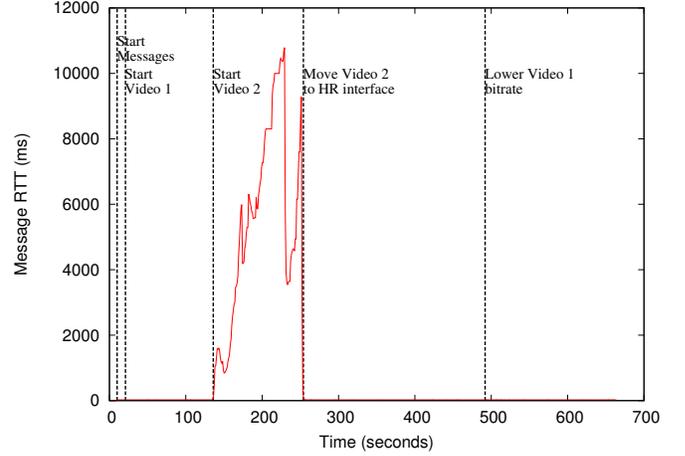

Fig. 5: SMS RTT for Scenario 1.

The tested scenario runs as follows:
1) Start the critical SMS server at t=0 seconds.
2) Initiate the high rate video server at t=0+20 seconds.
3) Initiate the low rate video server at t=0+150 seconds.

In our experiments, a manual MEC operator can, based on data gathered from the App. Manager, the SDN controller or the SDR controller, take the appropriate actions that guarantee the measurable requirements (Table I) and accept more applications at the same time (i.e fulfill more SLAs). Several actions can be carried out on the application flows. For example, a video flow can be moved to any of the interfaces, the high quality video (Video1) can be transcoded to a lower bitrate video etc. Our objective is to show that our framework makes available to MEC operators (automatized or manual), through specified interfaces, all the necessary information for an optimal configuration and management in a MEC context.

Figures 4 to 6 show the different actions taken in this scenario as well as the impact on different metrics. We voluntarily sequence the actions over time instead of combining them to highlight their individual impact. Figure 4 depicts the evolution of load on each link in time. Figure 5 shows the evolution of the RTT for SMS messages. Figure 6 plots the proportion of fulfilled SLAs.

As we start Video1, quite logically the load on the HR waveform increases. At this point, a single high bite rate video is being streamed on the HR radio and SMS are exchanged on the LR radio. The messages RTT is low and quite constant. We then activate Video2 on the LR interface through the App. manager and the SDN controller and see that the critical SMS RTT increases (Figure 5) as the load increases and reaches link capacity (around 125 Kbps in Figure 4). Noting the impact on the RTT and some available capacity on the HR link thanks to the SDR controller, the operator decides to move Video2 on the HR interface. This action, enforced by the SDN controller, is immediately followed by a drop in the RTT for SMS, but also impacts the load on the HR waveform, leading to saturation a few seconds later. To alleviate this situation, the operator decides to transcode Video1, thus lowering its bitrate.

[1] QoE refers to quality of experience of the observed video.

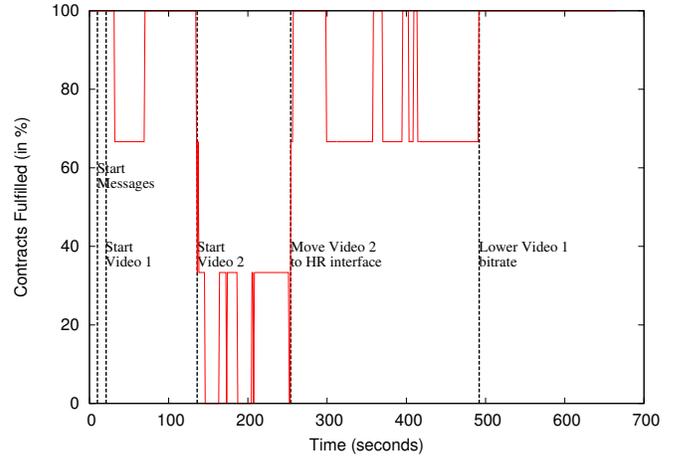

Fig. 6: SLAs satisfaction for Scenario 1.

This decision is implemented both by the App. Manager and the SDN controller upon request from the MEC controller which gathered information on the waveforms. This leads to a reasonable load (i.e far from capacity) on the HR interface. This action guarantees that all the requirements are met in our experiment as highlighted in Figure 6.

### C. Scenario 2: Guaranteeing the reliability of the LR waveform

In this scenario, the HR waveform has a capacity of 500 kbps while the LR interface has a capacity of 250 kbps. Only two applications are considered, Video1 and Video2 on the downlink, which can both be transcoded to lower their respective rate. We can also move the video flows from one interface to the other. Moreover, the code rate can be dynamically changed on the LR waveform to cope with random errors on the channel and increase its reliability. Finally, on the wireless links, our setup allows to introduce random errors following a uniform distribution with the mean of our choice. The SLAs are presented in Table II. At the beginning of the scenario, the code rate on the LR interface is equal to 1 (no code), we start both video flows.

TABLE II: Defined SLAs by application type (Scenario 2).

| Application | Requirements | Measured KPIs |
|---|---|---|
| Video1 | Good QoE | Bounded load on the HR waveform |
| Video2 | Good QoE | Bounded load and 95% success rate on the LR waveform |

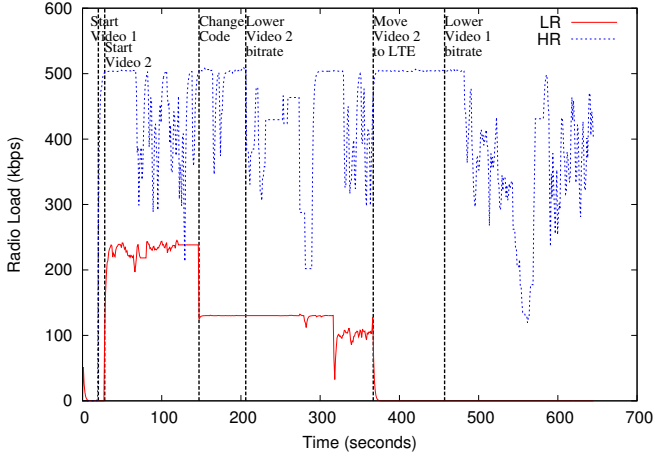

Fig. 7: Waveform loads for Scenario 2.

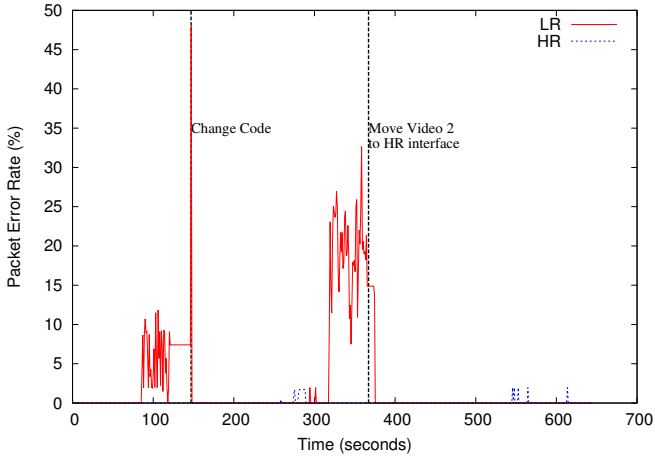

Fig. 8: Packet Error Rate (PER) for Scenario 2.

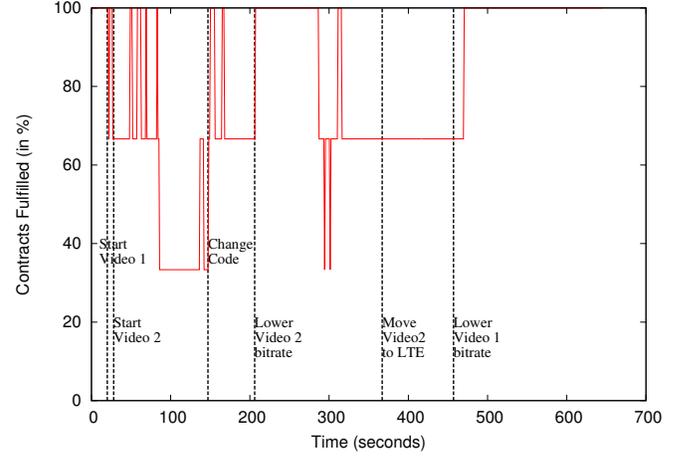

Fig. 9: SLAs satisfaction for Scenario 2.

a video transcoder through the App. manager and chaining the video server, the transcoder and the LR interface through the SDN controller.

At around t=0+300 seconds, errors on the LR link become high again (measures from the SDR controller). The operator thus decide to move the transcoded Video2 flow to the HR interface as Video2 cannot be further transcoded with acceptable QoE and therefore would not fit if the code rate is additionally reduced. This action is enforced by the SDN controller.

Moving Video2 to the HR interface leads to a saturation of the wireless link (refer to the load of HR in Figure 7 at t=0+350 seconds). Consequently, the operator then takes the action to transcode Video1 to a lower rate. This action, enforced by the App. manager and the SDN controller, alleviates the saturation problem and both video flows are transmitted on the HR interface with all SLAs finally satisfied.

## V. Conclusion and Perspectives

In this paper we have proposed a generic MEC framework called Multi-Radio Edge Node (MUREN) to address the challenge of using in real-time radio network information to improve service operation and thus user experience at the edge. Our work sets the basis and gives guidelines for combining SDR and SDN in MEC approach. We implemented a prototype of MUREN on a real testbed. The SDR part is based on USRP and GNU Radios while the SDN/NFV part is based on Open vSwitch and light containers. The experiments we conducted for two scenarios demonstrate the importance of taking the appropriate reaction at the right level (waveform, service chaining, or application). These reactions guarantee QoS requirement satisfaction for applications as well as an efficient use of the wireless resources.

In the future, we will work on decision algorithms that, based on monitoring information and available resources on the three layers (radio, network, application), will automatically optimize the delivery of services. Such algorithms aim at replacing the operator in the MEC controller of our architecture and thus automating the decision process. In addition, experiments with more nodes (UE) and different types of waveforms and services are also planned.

Figures 7 to 9 allow to follow closely the MEC network behavior as well as the operator actions. More precisely, Figure 7 shows the load on the different interfaces. Figure 8 shows the PER on the different interfaces. Figure 9 shows the proportion of fulfilled SLAs.

After the videos start, we observe that all application requirements are met until some errors occur on the LR link. In response to the errors, the MEC operator changes the code rate (to 1/2) on the LR interface at t=0+140 seconds. The measures are collected by the SDR controller. The action, enforced by the SDR controller, has two effects as it brings the measured error down to 0 but it also decreases the link capacity by a factor of 2, thus reducing the possible throughput to 125 kbps.

Indeed, the errors are down to 0 but Video2 is saturating the LR link (measures from the SDR controller). The MEC operator decides to transcode Video2 in order to reduce the load thus liberating some resources on the LR radio (Figure 7 at around t=0+205 seconds). This action consists of instantiating